\documentstyle[twocolumn,prl,aps,epsfig,amssymb]{revtex}
\tighten
\let\jnfont=\rm
\def\NPB#1,{{\jnfont Nucl.\ Phys.\ B }{\bf #1},}
\def\PLB#1,{{\jnfont Phys.\ Lett.\ B }{\bf #1},}
\def\EPJC#1,{{\jnfont Euro.\ Phys.\ J.\ C }{\bf #1},}
\def\PRD#1,{{\jnfont Phys.\ Rev.\ D }{\bf #1},}
\def\PRL#1,{{\jnfont Phys.\ Rev.\ Lett.\ }{\bf #1},}
\def\MPLA#1,{{\jnfont Mod.\ Phys.\ Lett.\ A }{\bf #1},}
\def\JPG#1,{{\jnfont J.\ Phys.\ G}{\bf #1},}
\def\CTP#1,{{\jnfont Commun.\ Theor.\ Phys.\ }{\bf #1},}

\begin{document}
\draft
\preprint{}
\title{Probing Topcolor-Assisted Technicolor \\from Top-Charm Associated
Production at LHC}

\author{Junjie Cao$^{a,b}$, Zhaohua Xiong $^c$, Jin Min Yang $^b$}

\address{$^a$ CCAST (World  Laboratory), P.O.Box 8730, Beijing 100080, China}
\address{$^b$ Institute of Theoretical Physics, Academia Sinica, Beijing 100080, China}
\address{$^c$ Graduate School of Science, Hiroshima University, Hiroshima 937-6256, Japan}
\date{\today}

\maketitle

\begin{abstract}
We propose to probe the topcolor-assisted technicolor (TC2) model
from the top-charm associated productions at the LHC, which are 
highly suppressed in the Standard Model. Due to the
flavor-changing couplings of the top quark with the scalars (top-pions
and top-Higgs) in TC2 model, the top-charm associated productions 
can occur via both the $s$-channel and $t$-channel parton processes 
by exchanging a scalar field at the LHC. We examined these processes 
through Monte Carlo simulation and found that they can reach the 
observable level at the LHC in quite a large part of the parameter 
space of the TC2 model.
\end{abstract}


The fancy idea of technicolor (TC) provides a possible mechanism
of breaking the electroweak symmetry dynamically. However, it is
hard for technicolor to generate the fermion masses, especially
the heavy top quark mass. As a realistic TC model, the
topcolor-assisted technicolor (TC2) model \cite{TC2} combines
technicolor with topcolor, with the former mainly responsible for
electroweak symmetry breaking and the latter for generating a
major part of top quark mass. This model is one of the promising 
candidates of new physics, awaiting to be tested at the upcoming 
CERN Large Hadron Collider (LHC). 

As shown in numerous previous studies, the top quark processes are
sensitive to new physics \cite{sensitive}. In some new physics
models like supersymmetry, there may emerge some new production
and decay mechanisms for the top quark at hadron colliders
\cite{new-production,new-decay,tcv_SUSY}. The study of these new
mechanisms will help to reveal the new physics effects.
It is noticeable that the TC2 model may have richer top-quark
phenomenology than other new physics models since it
treats the top quark differently from other quarks to single out
top quark for condensation \cite{TC2}. In fact, the top-color
interaction is flavor non-universal and consequently may induce 
new large flavor-changing (FC) interactions \cite{tb_TC2,gg2tc}.
One kind of such FC interactions occur between quarks and the top-pion, 
the pseudo-goldstone boson predicted by TC2 model with mass of a few 
hundred GeV, which are given by \cite{TC2,tb_TC2} 
\begin{eqnarray}
{\cal L}&=&\frac{m_t}{\sqrt{2} F_t} \frac{\sqrt{v^2-F_t^2}}{v}
\left[ i K_{UR}^{tt} K_{UL}^{tt \ast} \bar t_L t_R \pi^0_t   \right. \nonumber \\
&& \left. +\sqrt{2} K_{UR}^{tt \ast} K_{DL}^{bb} \bar t_R b_L
\pi^+_t + i K_{UR}^{tc} K_{UL}^{tt \ast} \bar t_L c_R \pi^0_t
\right. \nonumber \\
&& \left. + \sqrt{2} K_{UR}^{tc \ast} K_{DL}^{bb} \bar c_R b_L
\pi^+_t +h.c. \right], \label{coupl}
\end{eqnarray}
where $v \simeq 174 $ GeV, $F_t\simeq 50$ GeV is the top-pion
decay constant and the factor $\sqrt{v^2-F_t^2}/v$ reflects the
effect of the mixing between the top-pions and the would-be
goldstone bosons \cite{mixing}. $K_{UL}$, $K_{DL}$ and $K_{UR}$
are the rotation matrices that transform the weak eigenstates of 
left-handed up-type, down-type and right-handed up-type quarks to 
the mass eigenstates, respectively.
As pointed out in \cite{tb_TC2}, the
transition between $t_R $ and $c_R $, $K_{UR}^{tc} $, can be
naturally around $10\% \sim 30\%$ without conflict with low energy
experimental data and  $K$s can be parameterized as follows:
\begin{eqnarray}
K_{UL}^{tt}&=& K_{DL}^{bb}=1, \ \  \ \ K_{DR}^{tt}=1- \epsilon,
\nonumber \\
K_{UR}^{tc}&=& \sqrt{1-K_{UR}^{tt\ 2}-K_{UR}^{tu\ 2}} \leq \sqrt{2
\epsilon -\epsilon^2},
\end{eqnarray}
with $\epsilon$ representing the fraction of top quark mass
generated from TC interactions. Throughout this paper, we fix
$K_{UR}^{tc} =\sqrt{2 \epsilon -\epsilon^2} $ and treat $\epsilon
$ as an input parameter in the range $0 \sim 0.1 $. TC2 model also
predicts a CP-even scalar $h_t $ called top-Higgs\cite{gg2tc}. Its
couplings to quarks are similar to that of the neutral top-pion
except that the  neutral top-pion is CP-odd.
\begin{figure}[tbh]
\begin{center}
\vspace*{-2.0cm} \hspace*{-2cm} \epsfig{file=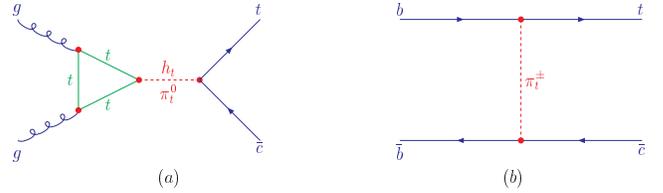,width=12cm}
\vspace*{-13cm} \caption{Feynman diagrams contributing to $t\bar
c$ associated production in TC2 model.} \label{fig1}
\end{center}
\end{figure}
The above large Yukawa couplings will induce the top-charm
associated production $pp\to t \bar c+X$ through both the
$s$-channel $gg\to \pi^0_t, h_t \to t \bar c$, as shown in
Fig.1(a), and the $t$-channel $bb \to t \bar c$ by exchanging a
$\pi^+_t$, as shown in Fig.1(b). Compared to the $t\bar b$
productions, which is also sensitive to TC2 \cite{tb_TC2}, the
$t\bar c$ productions are highly suppressed in the SM and thus its
observation would be a robust evidence of new physics.

The $t\bar c$ production through the $s$-channel $gg\to h_t \to t
\bar c$ has been briefly studied in the literature \cite{gg2tc}. But a
detailed Monte Carlo study of its observability, with the
consideration of the SM backgrouds, is still lacking. Given the 
great importance of the LHC phenomenology, such a study of observability
is absolutely necessary. Furthermore,
the $t\bar c$ production via the $t$-channel $bb \to t \bar c$,
which is more important when the top-pion is heavy,
has not been studied before. In this work we will give a
comparative study of both processes. We will not only study the
production rates, but also perform a Monte Carlo simulation to
show explicitly the observability at the LHC. Our result shows
that both the $s$-channel and $t$-channel processes can
reach the observable level at the LHC in quite a large part of 
the parameter space. Therefore, looking for the top-charm associated
productions at the LHC will serve as an important probe for the
TC2 model. 

The spin- and color-averaged amplitudes of the $s$-channel and
$t$-channel processes are given by
\begin{eqnarray}
|{\cal M}_s^{\pi^0_t}|^2&=& \frac{1}{64} \left ( \frac{m_t^2} {2
F_t^2} \frac{v^2 -F_t^2}{v^2} \frac{\alpha_s}{\pi} \right )^2 (2
\epsilon -\epsilon^2) (1-\epsilon)^2 \nonumber \\
&&\times \left\vert c_1 \left ( \frac{\hat{s}}{m_t^2} \right) \right\vert^2
\frac{m_t^2 (\hat{s} - m_t^2)}{(\hat{s}-m_{\pi^0}^2)^2 + m_{\pi^0}^2 \Gamma_{\pi^0_t}^2}, \\
|{\cal M}_s^{h_t}|^2&=& \frac{1}{64} \left ( \frac{m_t^2} {2
F_t^2} \frac{v^2 -F_t^2}{v^2} \frac{\alpha_s}{\pi} \right )^2 (2
\epsilon -\epsilon^2) (1-\epsilon)^2 \nonumber \\
&&\times \left\vert c_2 \left ( \frac{\hat{s}}{m_t^2} \right) \right\vert^2
\frac{m_t^2 (\hat{s} - m_t^2)}{(\hat{s}-m_{h_t}^2)^2 + m_{h_t}^2 \Gamma_{h_t}^2}, \\
|{\cal M}_t^{\pi^+_t}|^2&=& \left ( \frac{m_t^2} {2 F_t^2} \frac{v^2
-F_t^2}{v^2} \right )^2 (2 \epsilon -\epsilon^2)
(1-\epsilon)^2  \nonumber \\
&&\times \frac{\hat{t}(\hat{t}-m_t^2)}{(\hat{t}-m_{\pi^+}^2)^2},
\end{eqnarray}
where $\alpha_s =g_s^2/(4 \pi) $ with $g_s$ denoting strong
coupling constant, $\hat{s}$ and $ \hat{t} $ are patron level
Mandelstam variables. $\Gamma_{\pi^0_t} $ ($\Gamma_{h_t}$) is the 
width of neutral top-pion (top-higgs) which can be calculated by
considering all its decay modes. $c_{1,2}(R)$ are loop functions
defined by $c_1(R)=\int_0^1 {\rm d} x \frac{\ln{(1-R x (1-x))}}{x}
$ and $c_2(R)=-2+(1-\frac{4}{R}) c_1(R) $. Note there is no
interference between $M_s^{\pi^0_t} $ and $M_s^{h_t} $ due to
different CP property of $\pi^0_t $ and $h_t $. In our numerical
calculation, we use the CTEQ5L parton distribution
functions~\cite{pm} with $Q= \sqrt{\hat s}/2$.

\begin{figure}[b]
\begin{center}
\vspace*{-0.5cm} \epsfig{file=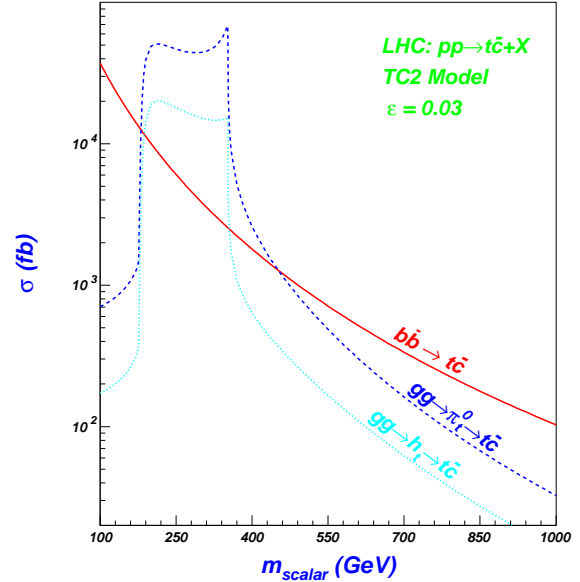,width=8cm} \vspace*{0.5cm}
\caption{Cross sections of top-charm associated production
as a function of corresponding scalar mass for different parton 
processes at the LHC.} \label{fig2}
\end{center}
\end{figure}

In Fig.\ref{fig2} we plotted the  two-body $t\bar c$ production
cross section versus corresponding scalar mass  for both the
$s$-channel and $t$-channel processes. The charge conjugate
production channel, i.e., the $\bar t c$ production, is also
included in our analysis throughout this paper.

We see from Fig.\ref{fig2} that for the  $t$-channel process the
production rate drops monotonously with the increase of the
charged top-pion mass. For the  $s$-channel processes the production
rates are maximum when the neutral scalars lie in the range of
$m_t +m_c \lesssim m_{\pi^0_t,h_t}\lesssim 2 m_t$. The reason is
that in this range, $t\bar{c}$ is the dominant decay mode of the
neutral scalars.  Comparing the rates of $s$- and $t$-channel productions, 
one sees that for a common scalar mass the $s$-channel rate is higher
than the $t$-channel rate only in the range from $m_t $ to $400$ GeV. 
From this figure one can also see that the cross section of
$p p \to \pi^0_t \to t \bar{c}$ is at least a couple of times larger 
than that of $p p \to h_t \to t \bar{c}$ for $m_{\pi^0_t} =m_{h_t} $.
Under the assumption that the top quark decays via the normal weak
interactions to $Wb$, the final state of  $t\bar c$ production is
$Wb\bar c$. We look for events with the leptonic decay of the
$W$, $W\to \ell \bar \nu$ ($\ell=e$ or $\mu$) and thus the signature
of $t\bar c$ production is an energetic charged lepton, one
$b$-quark jet, one light $c$-quark jet, plus missing $E_{T}$ from
the neutrino.  The potential SM backgrounds are the 
single top productions, top pair ($t\bar t$) productions, 
and $Wb\bar b$, $Wc\bar c$, $Wcj$ and $Wjj$  productions.
These backgrounds have been studied extensively in \cite{willen}.
In order to use the background results in  \cite{willen}, in our analysis
we applied the same selection cuts as in  \cite{willen}.    

First, we assumed silicon vertex tagging of the
$b$-quark jet with $60 \%$ efficiency and the probability of 0.5\%
(15\%) for a light quark ($c$-quark) jet to be mis-identified as a 
$b$-jet, which can reduce the background $Wjj$ efficiently.
We required the reconstructed top quark mass  $M(bW)$ to lie
within the mass range $\vert M(bW)-m_t \vert <20$ GeV,
which can reduce all the non-top backgrounds efficiently.

To simulate the detector acceptance,  we made a series of
basic cuts on the transverse momentum ($p_{T}$), the pseudo-rapidity
($\eta$), and the separation in the azimuthal angle-pseudo-rapidity plane
(~$\Delta R= \sqrt{(\Delta \phi)^2 + (\Delta \eta)^2}~ )$ between a jet and
a lepton or between two jets. These cuts are chosen to be
\begin{eqnarray}
& & p_T^{\ell}, ~p_T^j,~ p_T^{\rm miss} \ge  20 \rm{~GeV} ~,\\
& & \vert\eta_{\ell}\vert \le 2.5, ~\vert \eta_j\vert \le 4,
                                     ~\vert\eta_b\vert \le 2 ~,\\
& &\Delta R_{jj},~\Delta R_{j\ell} \ge 0.7 ~.
\end{eqnarray}
Further simulation of detector effects is made
by  assuming a Gaussian smearing of the energy of the final state
particles, given by
$\Delta E / E  =  30 \% / \sqrt{E} \oplus 1 \%$ for leptons
and $\Delta E / E =  80 \% / \sqrt{E} \oplus 5 \%$ for hadrons,
where $\oplus$ indicates that the energy dependent and independent
terms are added in quadrature and $E$ is in GeV.

Under the above cuts the total cross section of backgrounds is
16839 fb \cite{willen}, of which 7159 fb is from single top productions,
2770 fb from $t\bar t$ productions, and 5070 fb, 1460 fb, 
230 fb and 150 fb from $Wcj$, $Wjj$, $Wb\bar b$, $Wc\bar c$ productions,
respectively.

A few remarks are due regarding our analysis:

(1) The main kinematic difference between the $s$- and $t$-channel 
processes is that for the former the signal
has a peak in the top-charm invariant mass distribution
and this feature might be used to effectively reduce background
\cite{gg2tc}. In our analysis, we ignored this difference and adopt
the same strategy to probe the top-charm signal. The reason is twofold.
One is that without any information of $m_{\pi^0} $ and $ m_{h_t} $,
we do not know where the peak lies at. The other is, due to strong
interaction of the neutral scalars with quarks, their widths  are
of several hundred GeV for $ m_{\pi^0, h_t} > 2 m_t$ and as a
result, the peak is highly smeared.

(2) The two jets in the signal are 
$b\bar c$ (or $\bar b c$). Since the $c$-quark jet could be mis-identified 
as a $b$-jet with a probability of 15\%, the efficiency of tagging one
$b$-jet from $b\bar c$ (or $\bar b c$) should be slightly higher
than 60\%. In our analysis we conservatively assumed the tagging
efficiency of 60\%.

(3) For the same reason stated above, 
we could possibly require to tag two $b$-jets 
for the signal. Compared with tagging only one $b$-jet, this will further
reduce the signal rate by a factor of 15\% while suppress the $Wjj$ and $Wcj$ 
backgrounds by a factor of 0.5\%. However, the large backgrounds of top
quark productions 
cannot be relatively suppressed  because they contain two $b$-jets in their
final states. As a result, the total background is reduced only by a 
factor of 37\%. So this strategy of tagging two $b$-jets does no better.
 
\begin{figure}[b]
\hspace*{-1cm} \epsfig{file=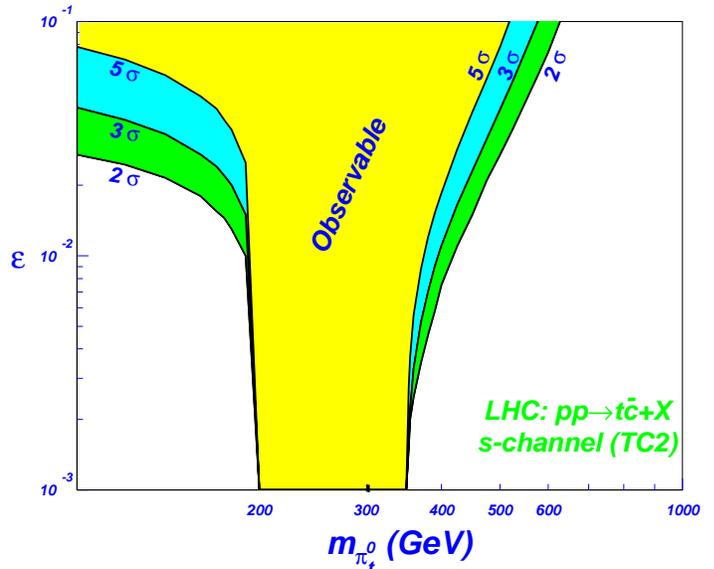,width=10cm} 
\vspace*{-1.2cm}
\caption{Observable region in the plane $(m_{\pi^0_t},\epsilon)$
for the  production $pp\to  t \bar c+X$
         through $s$-channel parton process $gg\to \pi^0_t \to t \bar c$  at the LHC with $L=100$ fb$^{-1}$.}
\label{fig-s}
\end{figure}

\begin{figure}[b]
\hspace*{-1cm} \epsfig{file=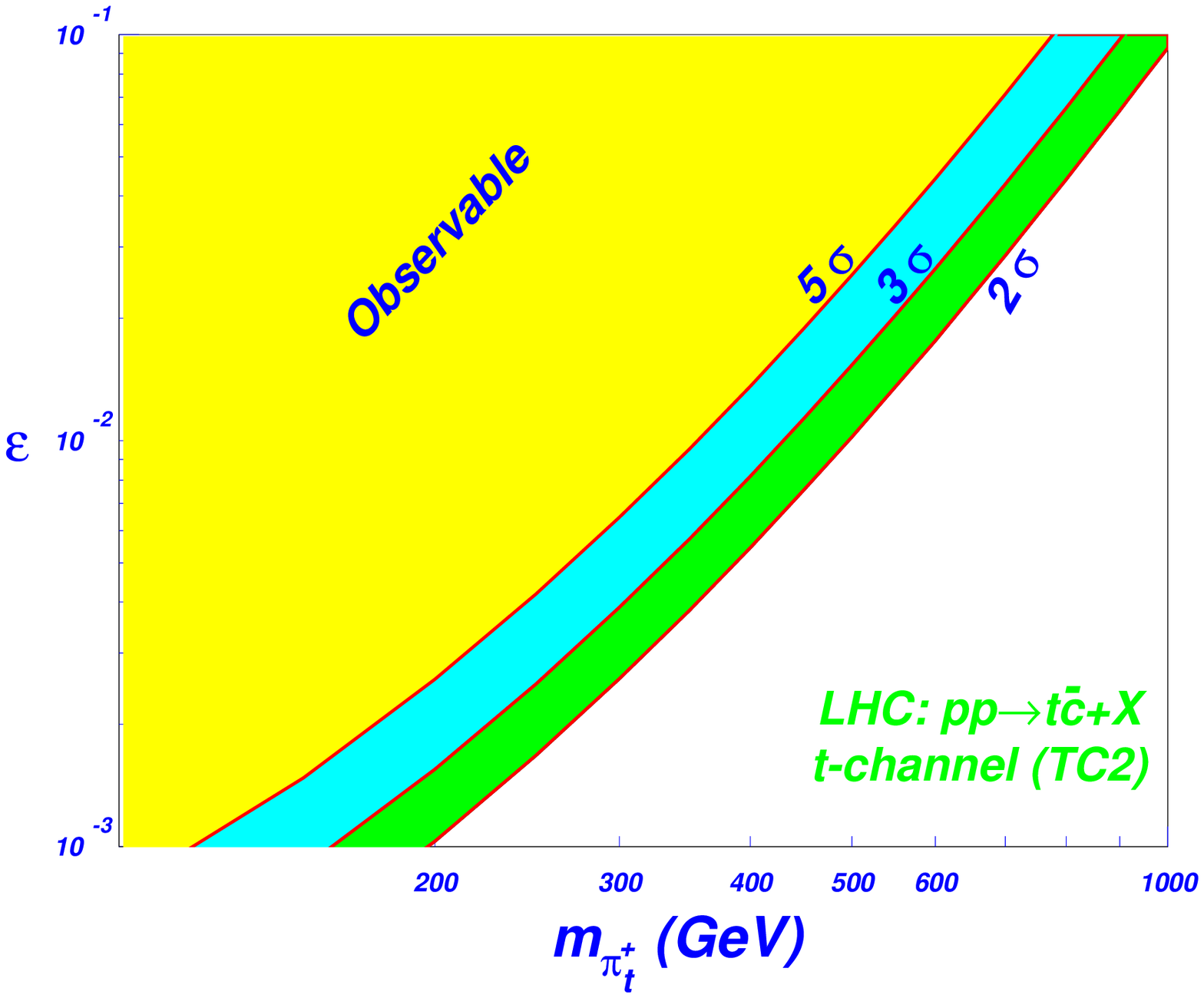,width=10cm} 
\vspace*{-1.2cm}
\caption{Same as Fif.\ref{fig-s}, but for the production $pp\to  t
\bar c+X$ through $t$-channel $b \bar b\to t \bar c$.}
\label{fig-t}
\end{figure}

The observability of the $t \bar c$ productions  through $g g \to
\pi_t^0 \to t \bar{c} $ and $ b \bar{b} \to t \bar{c} $ are shown
in Figs. \ref{fig-s} and \ref{fig-t} for the LHC with $L=100$
fb$^{-1}$. Throughout our analysis we restrict the value of the
parameter $\epsilon$ in the range of $0.001 \sim 0.1$.

We see from  Figs. \ref{fig-s} and \ref{fig-t} that for both
processes the observable parameter region is quite large. For the
$s$-channel process the region $m_t +m_c\lesssim
m_{\pi^0_t}\lesssim 2 m_t$ is observable for any $\epsilon$ value
in the range of $0.001 \sim 0.1$. The reason for this is already
elucidated in the discussions for Fig.\ref{fig2}. Outside the
region the signal is observable only for enough large $\epsilon$
value. But given  $0.001 \le \epsilon \le 0.1$ , the signal is
observable for $m_{\pi^0_t}\lesssim 550$ GeV. In case of
nonobservation, the $2\sigma$ lower limit on $m_{\pi^0_t}$ is about
$600$ GeV. 
\vspace*{-1.5cm}
\vspace*{-1.5cm}
 The $t$-channel process exhibits a different feature
from the $s$-channel process, which can be inferred from the
behavior of its two-body $t\bar c$ production rate shown in
Fig.\ref{fig2}. The observable region of the parameter space
shrinks monotonously with the increase of the charged top-pion
mass. But for a relatively high $\epsilon$ value in the range of
$0.001 \sim 0.1$, the signal is observable up to
$m_{\pi^+_t}\simeq 800$ GeV. This value is much larger than the
constraint from $R_b$\cite{rb}, {\it i.e.} $m_{\pi^+_t} \gtrsim
250 $ GeV. In case of nonobservation, the $2\sigma$ lower limit on
$m_{\pi^+_t}$ can reach $1$ TeV. 

We did not show the observable
region in the plane ($h_t $, $\epsilon$) for the s-channel 
process $ g g \to h_t \to t \bar{c} $ because it is similar to
Fig.\ref{fig-s}. The only difference is if $\epsilon \le 0.1 $,
the signal is observable at $3 \sigma $ level for $ h_t \lesssim
480 $ GeV and unobservable at $2 \sigma $ level for $h_t \gtrsim
530 $ GeV.

In our analysis of neutral scalar production, we did not choose $t
\bar{t} $ as the signal of the new physics for $m_{\pi^0}, m_{h_t}
> 2 m_t $. The reason is the effect of these scalars on the cross
section of  $t\bar{t} $ is  small, resulting in a contribution of
the order of $\sigma(t\bar{t})_{TC2}/\sigma(t \bar{t})_{SM}
=10^{-2} \sim 10^{-3}$ at the LHC.

In some extensions of the SM, such as the generic two-Higgs
doublet models with tree level FC scalar interactions \cite{2hdm},
both $s$- and $t$-channel top-charm associated productions can also
occur via exchanging neutral and charged Higgs bosons,
respectively. However, although the corresponding cross sections
are large, they are at least one order of magnitude smaller than
those of TC2 for a common mass of Higgs bosons and top-pions
\cite{tb_TC2}. The reason is that TC2 model predicts a large Yukawa
coupling $\frac{m_t}{2 F_t} \simeq 2.5$ and a possible large FC
coefficient $K_{UR}^{tc} $ (see Eq.(\ref{coupl})).

The interactions between the scalars and quarks can also induce
large neutral flavor changing gauge interaction, such as the
coupling $t\bar{c}g$ \cite{tcv_TC}. 
Since it is not easy to observe top rare
decay $t \to c g $ at hadron collider \cite{tcg},  such a coupling
may be well tested via single top production, $c g \to t $, at the
LHC. This process was extensively studied in effective Lagrangian
approach \cite{cgt} with the conclusion that for LHC with $100$
fb$^{-1}$ the discovery limit on the effective
interaction coefficient $\kappa_c/\Lambda $ is $0.0048$ TeV$^{-1}$ 
at $3 \sigma $ level, or alternatively, new physics contribution
to the cross section $ p p \to t^{\ast} \to b \bar{l}_e v_{e} $
must be larger than $40 $ fb. We examined this process in TC2 model
and found that for $\epsilon \le 0.1 $ only when $m_{h_t}\leq 400$
GeV and $m_{\pi^+, \pi^0} \leq 350$ GeV can such single top events be
observable. Obviously, the observable region is smaller than that
from top-charm associated production.

TC2 model also predicts some new top-quark decay modes, such as $t \to b
\bar{b} c $ via exchanging a $\pi^+_t $ and $t \to c WW $ via an
intermediate $h_t $. Among these new decay modes, $t \to b \bar{b}
c $ is dominant for a not heavy charged top-pion. For example, for
$\epsilon =0.1 $ and $m_{\pi^+} =250$ GeV, the ratio $\Gamma (t
\to b \bar{b} c )/\Gamma(t \to b W) $ can reach $9 \times 10^{-3}$,
which maybe not accessible at the LHC due to the SM backgrounds \cite{tch_mc}.
For the decay mode $t \to c W W$ \cite{tcww}, it drop quickly with 
the increase of $m_{h_t} $ and for $m_{h_t}>250$ GeV the ratio 
$\Gamma (t \to W W )/\Gamma(t \to b W) $ is of $10^{-7} $. As a result, 
such a decay mode cannot put severe bounds on TC2 parameter.

So we conclude the top-charm associated productions 
at the LHC are a powerful probe for the TC2 model.
The LHC can either observe the productions or set 
stringent bounds on the parameters of the TC2 model.

The work of Z.X. was supported by the postdoctor followship of
Japan Society for the Promotion of Science.  

\end{document}